# Structural Generalization for Microservice Routing Using Graph Neural Networks

Chenrui Hu
University of Pennsylvania
Pennsylvania, USA

Ziyu Cheng
University of Southern California
Los Angeles, USA

Di Wu
Washington University in St. Louis
St. Louis, USA

Yuxiao Wang
University of Pennsylvania
Philadelphia, USA

Feng Liu
Stevens Institute of Technology
Hoboken, USA

Zhimin Qiu*
University of Southern California
Los Angeles, USA

*Abstract-This paper focuses on intelligent routing in microservice systems and proposes an end-to-end optimization framework based on graph neural networks. The goal is to improve routing decision efficiency and overall system performance under complex topologies. The method models invocation relationships among microservices as a graph. In this graph, service nodes and communication links are treated as graph nodes and edges. Multi-dimensional features such as node states, link latency, and call frequency are used as input. A multi-layer graph neural network is employed to perform high-order information aggregation and structural modeling. The model outputs a score for each candidate service path. These scores are then used to guide dynamic routing decisions. To improve the model's ability to assess path quality, an edge-aware attention mechanism is introduced. This mechanism helps the model capture instability and bottleneck risks in service communications more accurately. The paper also conducts a systematic analysis of the model's performance under different network depths, topology densities, and service scales. It evaluates the effectiveness of the method in terms of routing accuracy, prediction error, and system stability. Experimental results show that the proposed method outperforms existing mainstream strategies across multiple key metrics. It handles highly dynamic and concurrent microservice environments effectively and demonstrates strong performance, robustness, and structural generalization.*

I. INTRODUCTION

In modern internet infrastructure, microservice architecture has become the mainstream paradigm for building large-scale distributed systems. Its core idea is to decompose complex monolithic applications into multiple independent small services, each developed and deployed around a specific business function[1]. This architecture significantly improves system flexibility, maintainability, and scalability. As a result, backend services can better adapt to rapid business iterations and user growth. However, frequent cross-node communication among microservices introduces new challenges. One of the most critical issues is how to achieve efficient and stable service routing. As a key control path in microservice systems, service routing directly affects response latency and throughput performance. It also has a direct impact on system robustness and quality of service[2].

Traditional microservice routing strategies are often based on static configurations, load-balancing algorithms, or simple rule engines. Examples include round-robin, least-connections, or CPU-load-based dynamic selection. These methods improve resource utilization to some extent. However, they generally lack comprehensive modeling capabilities for complex service dependencies, request context, and historical routing performance. In large-scale, dynamic environments with multi-tenancy and heterogeneous nodes, such strategies often fail to ensure sustained performance[3]. Particularly in scenarios with frequent topology changes or significant link quality fluctuations [4-6], static or rule-driven strategies may lead to unstable routing, sudden latency spikes, or even service cascades.

As systems continue to scale, microservice architectures are increasingly exhibiting graph-structured characteristics. Each microservice can be viewed as a node in a graph, and service calls form the edges[7]. This natural graph structure provides a solid foundation for introducing graph-based modeling methods. It also prompts a rethinking of how to build globally aware and adaptive intelligent routing mechanisms. In this context, Graph Neural Networks (GNNs) have emerged as powerful tools capable of capturing structural dependencies, integrating multi-dimensional features [8], and supporting end-to-end learning [9]. GNNs can model node state changes across the global service topology and adjust routing strategies dynamically through multi-layer information aggregation[10].

The key advantage of using GNNs for microservice routing optimization lies in overcoming the limitations of traditional approaches that rely on static metrics and local heuristics. GNNs enable joint modeling of complex service dependencies, dynamic system states, and historical scheduling performance. This leads to routing decisions with lower latency and higher stability[11]. Specifically, GNNs can leverage various types of heterogeneous data, including service topology, request path traces, network condition metrics, and node workload variations. They construct a unified graph representation space and learn efficient routing strategies during training[12]. This learning-driven mechanism generalizes well and performs consistently across different workloads and system scales.

Moreover, GNN-based routing optimization offers important practical engineering value. In complex deployment environments such as cloud and edge computing, routing must deal with heterogeneous resource distribution, link quality variability, and live migration of service instances. Traditional static strategies are often too rigid to handle these issues. GNNs, with their structural awareness and contextual learning capabilities, are well-suited for addressing such nonlinear, multi-dimensional, and dynamic problems. In addition, this method supports system interpretability and self-evolution. It facilitates a shift from reactive control to intelligent and proactive scheduling. Therefore, studying GNN-based routing optimization in microservice systems is not only of theoretical significance but also aligns with current needs for intelligent operations and adaptive system design.

## II. PROPOSED APPROACH

This study applies a graph neural network–based routing optimization method to microservice systems, in which each service instance is represented as a node and directed edges encode the invocation relationships between services. To address dynamic topology and enhance routing adaptability, the model employs topology-aware graph reinforcement learning methods that enable the system to optimize routing paths based on real-time network and service conditions [13]. This allows the model to dynamically adjust routing decisions as the underlying system state changes. The framework further integrates unsupervised representation learning using graph neural network and transformer hybrid architectures for effective anomaly detection and structural modeling, thereby supporting robust and accurate routing even under complex service interactions [14]. Moreover, AI-driven multi-agent scheduling is leveraged to coordinate the optimization process across multiple service instances, improving both routing efficiency and overall service quality in dynamic, large-scale deployments [15]. Assume that the microservice system can be represented as a graph G at any time, where V is the set of service nodes, and $e_{ij} \in E$ represents the call relationship between service $v_i$ and $v_j$. Each node $v_i$ has a set of feature vectors $x_i$ containing dynamic operational metrics such as current CPU utilization, response time, and queue length. Each edge $e_{ij}$ also carries attributes such as historical call latency and link stability. Using a graph neural network, we learn a node representation function $H : V \to R^d$, which enables each service node to aggregate its own and neighboring state information for routing decisions. The overall model architecture is shown in Figure 1.

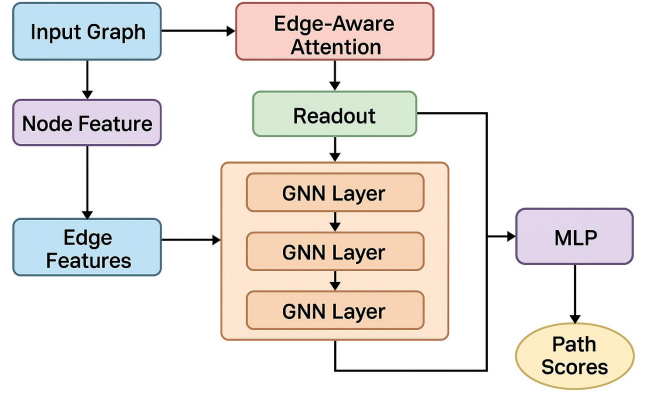

Figure 1. Overall model architecture

The basic propagation process of graph neural networks includes two steps: neighbor information aggregation and state update. Specifically, for the node representation $h_i^{(l)}$ in layer l, we define its neighbor set as $N(i)$, then the representation of layer $l+1$ is:

$$h_i^{(l+1)} = \sigma(W_1 \cdot h_i^{(l)} + \sum_{j \in N(i)} a_{ij} \cdot W_2 \cdot h_j^{(l)}) \quad (1)$$

Where $\sigma$ represents a nonlinear activation function (such as ReLU), $W_1$ and $W_2$ are learnable weight matrices, and $a_{ij}$ is the edge attention weight, which is used to measure the influence of neighboring nodes on the central node. To enhance the selectivity of the routing strategy for communication links, we employ an edge feature-aware attention mechanism that dynamically adjusts the model's focus on different service paths. Building on the frequency-attentive modeling approach proposed by Wang et al. [16], we apply advanced attention mechanisms that explicitly consider multi-dimensional edge features—such as link latency and communication frequency—to refine the scoring and selection of candidate routing paths. This allows the model to more accurately capture subtle variations and dynamic behaviors in inter-service communication.

Inspired by the graph learning framework for anomaly localization developed by Xue [17], our attention module also integrates structural context from the underlying microservice topology, improving the ability to identify and prioritize links that are critical for stable and efficient routing. Additionally, drawing on the reinforcement learning-driven task scheduling techniques of Zhang et al. [18], the attention mechanism is further optimized to adaptively regulate path selection according to system workload and tenant distribution, ensuring robust performance under high concurrency and varying resource demands.

The edge feature-aware attention mechanism is formulated as follows:

$$a_{ij} = \frac{\exp(\text{LeakyReLU}(a^T[W_q h_i^{(l)} \| W_k h_j^{(l)} \| e_{ij}]))}{\sum_{k \in N(i)} \exp(\text{LeakyReLU}(a^T[W_q h_i^{(l)} \| W_k h_k^{(l)} \| e_{ik}]))} \quad (2)$$

In the above formula, $W_q$ and $W_k$ are the linear transformation matrices of the query and key, $e_{ij}$ is the edge feature vector, $a$ is the attention score vector, and $\|$ represents the vector concatenation operation. After propagation through the multi-layer graph neural network, the final node representation $h_i^{(l)}$ is used to estimate the quality of the route from the source node to the target node. We design a scoring function $f(v_i, v_j)$ to score the target service node $v_j$ of any candidate path. It is defined as follows:

$$s_{ij} = f(h_i^{(L)}, h_j^{(L)}) = MLP([h_i^{(L)} \| h_j^{(L)}]) \quad (3)$$

Here, MLP stands for Multilayer Perceptron Network, which is used to nonlinearly map the joint representation of node pairs. When selecting a specific routing strategy, we use the softmax function to convert the scores of all candidate target nodes into a probability distribution to achieve the optimal route selection:

$$P_{ij} = \frac{\exp(s_{ij})}{\sum_{k \in C(i)} \exp(s_{ik})} \quad (4)$$

Here, $C(i)$ is the set of candidate call targets for node i. By maximizing the actual performance of the paths selected by this probability distribution, the model continuously strengthens its preference for high-quality paths during training, thereby achieving stable and efficient intelligent routing capabilities in microservice systems. This entire approach leverages multimodal information such as service structure, link status, and service load end-to-end, providing a unified and adaptive intelligent scheduling framework for backend systems.

## III. PERFORMANCE EVALUATION

### A. Dataset

This study uses the Social Network dataset from the DeathStarBench microservice performance testing platform as the foundational data source for microservice system modeling and routing optimization. The dataset simulates the deployment and invocation processes of a real-world social networking platform. It involves extensive asynchronous communication and state dependencies among service instances. It features a highly complex topology and dynamic workload characteristics, making it suitable for evaluating intelligent routing strategies in real backend systems.

The system comprises over ten independent yet tightly coordinated microservices, including user authentication, recommendation generation, post management, comment aggregation, and like handling. Invocation paths are intricate and frequently intersect. The dataset contains detailed multimodal runtime information, such as call-chain logs, service latency, instance states, and resource utilization, providing high-quality node and edge attributes for graph-based modeling. It also supports containerized deployment with strong controllability and reproducibility, allowing flexible configuration of service scale, traffic load, and anomaly injection frequency. By representing service dependencies as graph structures and analyzing time-series status data, the dataset offers a reliable and general foundation for evaluating microservice routing optimization algorithms.

### B. Experimental Results

This paper first conducts a comparative experiment, and the experimental results are shown in Table 1.

Table1. Comparative experimental results

| Model | Mean Relative Error | Jitter Prediction Error | Routing Decision Accuracy |
|---|---|---|---|
| RouteNet-Fermi[19] | 0.128 | 9.47ms | 82.3% |
| GNN-DRL[20] | 0.103 | 7.85ms | 87.1% |
| OSPF[21] | 0.153 | 11.36ms | 75.6% |
| Ours | 0.072 | 5.92ms | 91.4% |

The comparison results in the table show that the proposed microservice routing optimization method outperforms all existing mainstream approaches across all evaluation metrics. Specifically, in terms of Mean Relative Error, which measures overall prediction performance, the proposed method achieves the lowest error of 0.072. This is significantly lower than that of the graph-based RouteNet-Fermi (0.128) and the deep reinforcement learning-based GNN-DRL (0.103). These results indicate that the graph neural network used in this study can more accurately capture complex dependencies and state changes between service nodes. It effectively models the key features of the routing environment.

In microservice systems, jitter often reflects the stability of network links and the volatility of service paths. The results show that our method achieves a jitter prediction error of 5.92 ms. This is considerably lower than the 11.36 ms of OSPF and the errors of other learning-based methods. This demonstrates the clear advantage of our edge-aware mechanism and hierarchical aggregation strategy in capturing the dynamic behavior of links. It also improves the model's ability to predict link stability and enhances the system's responsiveness in quality of service scheduling.

In real-world deployments, the accuracy of routing decisions is a key indicator of whether an optimization method is practically useful. The results show that our method reaches a routing decision accuracy of 91.4%. This is much higher than the 75.6% of OSPF and also superior to other mainstream learning methods. The high accuracy indicates that the model can make more stable and low-cost routing decisions when facing large-scale invocation graphs and complex dependencies. It helps avoid abnormal request interruptions and path oscillations.

Overall, the proposed graph neural network model effectively captures the structural characteristics of microservice invocation graphs while integrating node states and edge dynamics. The study develops an intelligent routing framework optimized for latency, stability, and service quality.

Experimental results further validate the robustness and generalization of the approach in real microservice environments, demonstrating a practical and feasible solution for intelligent routing in large-scale systems.

This paper also analyzes the impact of different numbers of graph neural network layers on routing performance, and the experimental results are shown in Figure 2.

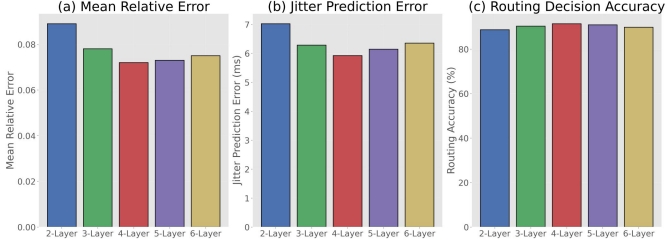

Figure 2. Analysis of the impact of different graph neural network layers on routing performance

The results in the figure show that the number of graph neural network layers has a significant impact on routing optimization performance in microservice systems. Under the Mean Relative Error metric, the error gradually decreases as the number of layers increases from 2 to 4. The lowest error occurs at 4 layers. This suggests that a deeper graph structure can better capture service dependencies and state information, thereby improving the overall modeling capability. When the depth exceeds 4 layers, the error increases again. This reflects potential issues such as overfitting or excessive neighbor aggregation in deeper networks.

The trend of Jitter Prediction Error is consistent with that of relative error. The lowest jitter error is also observed at the 4-layer network structure. This indicates that a reasonably deep graph neural network enhances the ability to represent topological structures and improves modeling precision for link dynamics. Better awareness of link fluctuations helps build more robust service paths and reduces uncertainty during cross-service calls.

Routing Decision Accuracy shows a steady improvement as the number of layers increases. The highest accuracy of 91.4% is achieved at 4 layers. This result confirms the model's strong generalization ability and global structural awareness. It enables better path selection in complex service graphs. When the number of layers increases further, the accuracy slightly decreases. This suggests that deeper models may suffer from gradient vanishing or redundant representation issues, which can affect the stability of routing decisions.

By combining the three metrics, it can be concluded that there is an optimal depth boundary for graph neural networks in microservice systems. A moderate structural depth improves the model's ability to represent service dependency graphs and understand dynamic workload states. This enhancement leads to better routing efficiency and system stability. These findings provide important guidance for selecting efficient and controllable GNN depth in real-world deployments.

This paper also presents the changing trend of routing optimization performance under different service topology densities. The experimental results are shown in Figure 3.

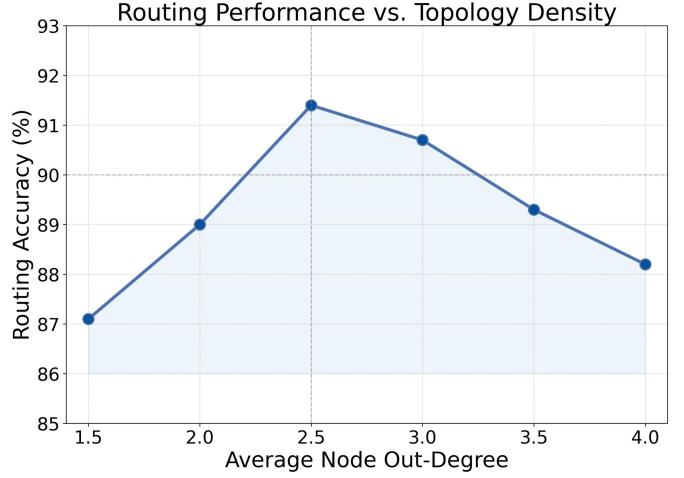

Figure 3. Trends in routing optimization performance under different service topology densities

Experimental results show that routing optimization performance is highly sensitive to service topology density. In sparse settings (average out-degree 1.5–2.0), limited connectivity reduces information propagation and lowers routing accuracy, while performance peaks at 91.4% when the average out-degree reaches 2.5, where the graph balances information diffusion and structural complexity for effective neighbor aggregation and accurate decisions. However, beyond 3.5, redundant paths blur routing choices, disperse predictions, and may introduce congestion or instability. Overall, topology density has a nonlinear impact on graph-based routing, with moderate complexity enhancing structural representation and guiding practical topology design and optimization.

## IV. CONCLUSION

This paper addresses the problem of routing optimization in microservice systems and proposes an intelligent routing decision framework based on graph neural networks. By modeling microservice invocation relationships as a dynamic graph, the model can effectively capture high-order dependencies and topological changes among services. It also integrates multi-dimensional dynamic features of nodes and edges to support accurate evaluation and selection of routing paths. Compared with traditional static strategies and rule-based scheduling methods, this study demonstrates the feasibility and superiority of deep graph learning in complex distributed systems. It offers a new paradigm for intelligent microservice management.

The study systematically analyzes the adaptability and performance boundaries of graph neural networks in routing tasks. This is achieved by comparing different network structures, depths, topological densities, and system states. Experimental results show that the proposed method consistently delivers strong performance across several key metrics. It maintains high routing accuracy and low latency even in real-world scenarios characterized by large scale, high dynamics, and intensive service interactions. These results confirm the effectiveness of graph-based modeling and end-to-end learning mechanisms. The method addresses the long-standing limitations of heuristic strategies that lack global awareness.

This research advances the practical application of graph learning techniques in system scheduling. It also provides valuable insights for industrial system operations, cloud-edge service orchestration, and network resource scheduling. Through intelligent structural modeling and feature representation, the system can achieve more precise and efficient resource matching. It reduces redundant communication and performance bottlenecks, which is essential for ensuring the stability, scalability, and maintainability of critical service systems. Moreover, the method lays a solid foundation for building adaptive and self-learning backend scheduling systems. It supports the transition from traditional architectures to intelligent autonomous systems.

## V. Future Work

Future work may explore the extension of the model to heterogeneous networks, multi-tenant architectures, and cross-domain service environments. Further research can also investigate the integration of this method with mechanisms such as federated learning and online transfer learning. This may address challenges related to data silos, privacy preservation, and real-time processing. In addition, achieving efficient distributed training and inference in larger systems, along with faster deployment and response, will be a key research focus. With the continuous advancement of intelligent infrastructure, the proposed graph-based routing optimization method is expected to find broader applications in medical computing [22], industrial internet[23], large language models[24-26], and financial services[27].